# Formation mechanism of the (2 × 1) reconstruction of calcite (104)

Haojun Zhou[1], Yingquan Chen[1], Mingyue Ding[2] and Xiaoliang Zhong[1,*]

[1]School of Energy and Power Engineering, Huazhong University of Science and Technology, Wuhan 430074, China
[2]School of Power and Mechanical Engineering, Wuhan University, Wuhan 430072, China.

e-mail: xzhong@hust.edu.cn

## Abstract

Calcite has recently attracted extensive research interest in fields ranging from geoscience to carbon dioxide removal. Although much effort has been made to study the (2x1) reconstruction of the most stable (104) surface, the origin of this reconstruction remains unclear. Here, we carefully investigate the atomic and electronic structures of calcite (104) via density functional theory methods with van der Waals corrections. The results unambiguously show that the driving force for this reconstruction is the intrinsic demands of surface atoms to increase the coordination numbers. On reconstructing, calcite (104) forms four additional Ca-O bonds per (2x1) unit cell. Besides, phonon spectrums indicate both unreconstructed and reconstructed surfaces are dynamically stable. Finally, by applying the climbing image nudged elastic band method, an energy barrier is predicted during the reconstructing. This work delivers a full picture for the formation of calcite (104)-(2x1) reconstruction and can greatly advance the understanding of surface science for calcite.

Calcium carbonate ($CaCO_3$) is one common substance on Earth. It is the major constituent of limestone, marble, eggshells, and pearls. As the major polymorph of $CaCO_3$ in nature[1], calcite has been actively investigated in fields including geoscience[2, 3, 4], soil stabilization[5, 6], carbon dioxide removal[7], and new material development[8, 9]. The most stable face of calcite is the (104) surface with the fewest Ca–O bonds broken upon surface formation[10]. This surface supports virtually all processes involving calcite[11]. It is known that surface reconstruction can have a profound impact on crystal growth, surface properties and application of crystalline materials[12, 13, 14]. Since 1990s, a range of surface sensitive analysis techniques including low-energy electron diffraction (LEED)[15], X-ray photoelectron spectroscopy (XPS)[15, 16], grazing incidence X-ray diffraction (GIXRD)[17] and atomic force microscopy (AFM)[11, 18, 19, 20, 21] have been applied to detect the (2x1) reconstruction of calcite (104). On the other hand, theoretical studies including molecular dynamics (MD) simulations[10, 22, 23] and density functional theory (DFT) calculations[11, 21, 24] have been performed to search for the atomic model of this reconstructed surface and to provide insight into the origin of the (2x1) reconstruction. Recently, by combining high-resolution non-contact atomic force microscopy (NC-AFM) data and DFT calculations, it is established that the (2x1) reconstruction is thermodynamically the most stable form of calcite (104)[11].

Regarding the origin of the calcite (104)-(2x1) reconstruction, the early views proposed in the 1990s include cation ordering[25, 26]. This viewpoint was recently rebutted by Rohl et al. via DFT calculations[11]. In 2003 Rohl et al. used MD methods and pointed out that an imaginary phonon mode of the unreconstructed surface signifying the necessity of surface reconstruction[22]. Nevertheless, vibrational properties obtained by MD simulations should be treated with caution since these simulations use empirical potentials[27] and it is known that in some cases MD methods don't yield satisfactory results for calcite[28]. Calculations based on more accurate quantum mechanical methods are thus highly desired to check this. Stipp et al. also performed MD simulations and proposed that step edges may be responsible for the (2x1) reconstruction[10]. However, the step edge effect becomes negligible for terrace sizes larger than 4.5 nm[10] and subsequent AFM studies where cleaved surfaces were shown to be flat reported the reconstruction[11, 19]. Very recently, Rahe et al. stated that the origin of the (2x1) reconstruction was 'purely thermodynamic' via DFT simulations[11]. Yet it remains unclear why the reconstructed surface has a lower total energy. Besides, it is not known whether the (2x1) reconstructing is a spontaneous process, or an energy barrier exists during the reconstructing, i.e., the nature of reconstruction.

In the present work, we make efforts towards obtaining a fundamental understanding of the calcite (104)-(2x1) reconstruction by applying DFT methods. We note that DFT simulations have recently played an important role in elucidating the mechanisms of crystal surface reconstruction[29, 30, 31, 32] due to the relatively high accuracy and capability of investigating the underlying electronic structure in depth. Our results unambiguously show that the (2x1) reconstruction is resulted from the intrinsic demands of surface atoms to increase the coordination numbers. We have also calculated vibrational properties to check if the unreconstructed surface has imaginary phonon modes. Finally, the climbing image nudged elastic band (CI-NEB) method[33] is applied to ascertain if there is an energy barrier during the reconstructing.

## Results and discussion
### Structure characteristics of reconstructed calcite (104)

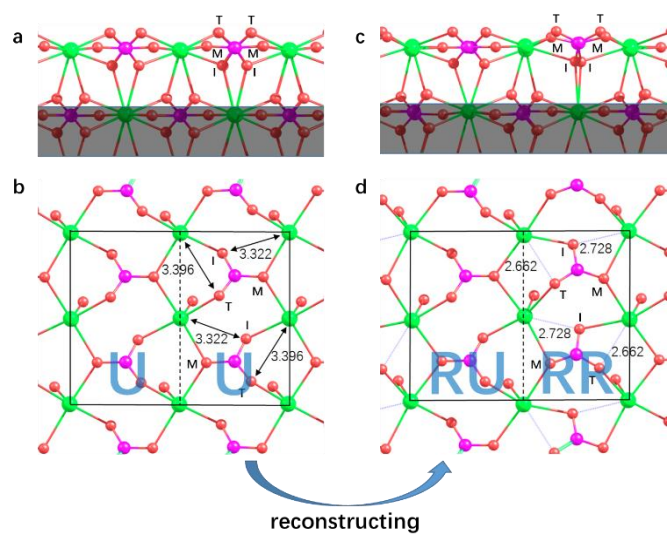

**Fig. 1 | Atomic structures of unreconstructed and reconstructed calcite (104).** Top and bottom panels are side view and top view, respectively. The shaded area of the side view is not shown in the top view for clarity. The unreconstructed surface (U) is shown on the left while the reconstructed one is shown on the right. RU and RR represent the unreconstructed half and the reconstructed half of a (2x1) unit cell of the reconstructed surface, respectively. Colour scheme: Ca in green, C in purple, and O in red. T = topmost, M = middle, I = innermost.

We first describe the structural features of both unreconstructed and reconstructed calcite (104) surfaces. The optimized calcite bulk lattice constants are a = b = 5.027 Å, and c = 17.061 Å, compared well with the experimental values[34] (4.991 and 17.062 Å). In calcite bulk, each Ca atom is bonded to six O atoms while each O atom binds with two Ca atoms. All Ca-O bond lengths (nearest Ca-O distance) are 2.368 Å. The next-nearest (NN) Ca-O distance is considerably greater (3.481 Å). In the top layer of the unreconstructed surface, coordination number (CN) of each Ca atom is reduced to five (Fig. 1a, b). Each outmost O atom of one carbonate group binds with only one Ca atom while the other two O atoms still bind with two Ca atoms. The (2x1) reconstruction features a rotation of the carbonate group within one half (called 'the reconstructed half', denoted by '**RR**' hereafter) of each (2x1) unit cell while structure variation in the other half (called 'the unreconstructed half', denoted by **RU**) is much less apparent[11, 22] (Fig. 1c,d). Atomic structure of the optimized reconstructed (104) surface is essentially the same as that obtained by Rahe[11] since both simulations are at the same level of theory. It is to be noted that based on this structure, NC-AFM images at different tip-sample distances have been successively reproduced[11]. Previously, a similar reconstructed structure obtained by MD simulations was shown to be able to reproduce the experimental LEED pattern[22].

We identify that upon surface reconstructing, each surface $CO_3^{2-}$ group in RR forms two *additional* Ca-O bonds. As Fig.1 shows, each topmost O atom (T) in RR moves towards one NN Ca atom. Rotation of the $CO_3^{2-}$ group of about 18° renders a significant decrease in the corresponding Ca-O distance from 3.396 Å (Fig. 1b) to 2.662 Å (Fig. 1d), i.e., a reduction of about 0.7 Å. Likewise, distance between each innermost O atom (I) and one NN Ca atom substantially decreases from

3.322 Å to 2.728 Å (reduction ≈ 0.6 Å). The two Ca-O bonds (2.662 and 2.728 Å) formed upon reconstruction are about 0.3 Å longer than the nearest Ca-O distance in the bulk (2.368 Å) and are about 0.8 Å shorter than the NN Ca-O distance (3.481 Å). One may raise a question why previous studies haven't discussed these two additional bonds (as far as the authors are aware). One reason may be that these chemical bonds are relatively longer such that some graphical programs don't show the bonding, therefore the drastic changes in Ca-O distances were overlooked. On the other hand, all five nearest Ca-O distances associated with one $CO_3^{2-}$ group within the right half of the (2x1) cell in Fig. 1b increase upon reconstructing, but the magnitude of variation is relatively small (no greater than 0.11 Å). In the left half of the (2x1) cell in Fig. 1b, three out of the five nearest Ca-O distances associated with one $CO_3^{2-}$ group increase upon reconstructing, with maximum change being ~0.05 Å. The other two nearest Ca-O distances decrease slightly (change no greater than ~0.01 Å).

**Electronic structure analyses on calcite (104)**

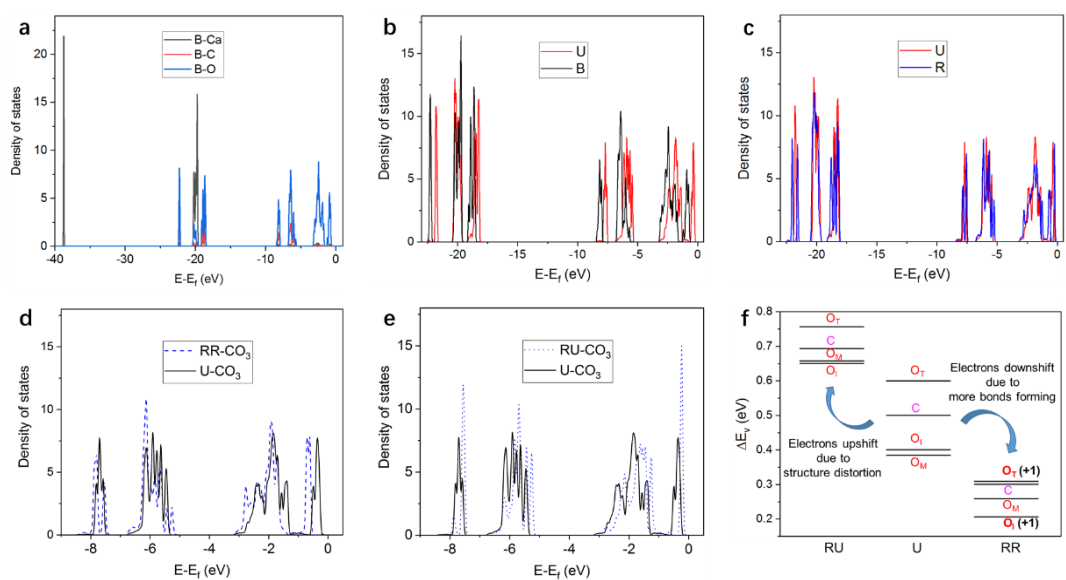

**Fig. 2 | Calcite electronic structures: bulk, unreconstructed and reconstructed (104). a** DOS of calcite bulk (B) projected on different species. **b** DOS of bulk and unreconstructed (U) (104). **c** DOS of unreconstructed and reconstructed (R) (104). **d** DOS associated with one carbonate group in RR (RR-CO3) versus in the unreconstructed surface (U-CO3). **e** DOS associated with one carbonate group in RU (RU-CO3) versus in the unreconstructed surface. **f** Shift of valence electron levels associated with one carbonate group in U, RU and RR. $O_T$, $O_I$ and $O_M$ stand for the topmost, innermost and middle oxygen atoms, respectively (see Fig. 1). '+1' in the brackets means coordination number increases by 1.

We now discuss calcite electronic structures which underlie the energetics. Fig 2.a shows the density of states (DOS) of calcite bulk. In line with literature[35], Ca shows high peaks around -38 and -20 eV, respectively. DOS is dominated by C and O electrons in the energy range close to Fermi level (-10 to 0 eV). Moreover, O has much higher DOS than C does, as a result of the 3:1 atomic ratio and the stronger ability to attract electrons. Atomic charge is estimated by Bader analysis. In calcite bulk, atomic charge of Ca, C and O is +1.63, +2.17 and -1.27, respectively. Due to the reduced coordination numbers, energies of surface atoms are higher than those of the bulk counterparts. Accordingly, DOS spectrum of the unreconstructed surface in general shifts to the right compared with that of the bulk (Fig 2.b). For clarity surface DOS peak around -38 eV is not shown in Fig 2.b, which only differs from the corresponding bulk peak slightly. On the other hand, change in atomic charge is small at the surface, with maximum variation smaller than 0.02 for all three species. DOS of the surface is taken to be that associated with atoms in the top layer. We show in Supplementary Fig. 1 that DOS of inner layers only differs from that of bulk slightly. Fig 2.c compares DOS spectra of unreconstructed and reconstructed surfaces. One notes that DOS peaks of the reconstructed surface are typically lower and broader, reflecting the lower symmetry of the reconstructed structure. Nevertheless, DOS spectra of both surfaces in general overlap with each other, underlying the nearly degenerate surface energies[11, 24].

We show in Fig. 1 that on reconstructing, the surface forms four additional bonds per (2x1) unit cell. We now discuss the impact of these 'newly formed' bonds on electronic structures. Fig. 2 d shows that DOS spectrum associated with each $CO_3^{2-}$ group in RR shifts to the lower energy range compared with that in U. DOS below -9 eV is not shown for clarity, which also shifts to the lower energy range. This means $CO_3^{2-}$ groups in RR are thermodynamically more stable than those in U. In contrast, DOS spectrum of each $CO_3^{2-}$ group in RU shifts to the higher energy range (Fig.2 e). To better analyze the shift in DOS, we calculate the (average) valence electron energy level $E_{v-x}$ of an atom $x$ by Eq. (1)

$$E_{v-x} = \frac{\int_{-\infty}^{E_f} E_x D(E_x) dE_x}{\int_{-\infty}^{E_f} D(E_x) dE_x} \qquad (1)$$

where $E_x$ is valence electron energy of atom $x$ and $D(E_x)$ is the corresponding DOS. We then define valence electron level shift ($\Delta E_{v-x}$) to be the difference between $E_{v-x}$ of a surface atom and that of its bulk counterpart as given by Eq. (2)

$$\Delta E_{v-x} = E_{v-x}(\text{surface}) - E_{v-x}(\text{bulk}) \qquad (2)$$

As Fig. 2f shows, at the unreconstructed surface the topmost oxygen atoms ($O_T$) have the highest $\Delta E_v$ as a result of CN reduction (from two to one). Upon reconstructing, in RR each outmost oxygen atom ($O_T$) and each innermost oxygen atom ($O_I$) forms one additional bonds with Ca (Fig. 1). $O_T$ and $O_I$ in RR exhibit noticeable lowering in valence electron energy, with change in $\Delta E_v$ being -0.29 and -0.19 eV, respectively (Fig. 2f). As a result, the difference between $\Delta E_{v-O_T}$ and $\Delta E_{v-C}$ is much smaller in RR than in U. Regarding $O_I$, in RR $\Delta E_{v-O_I}$ becomes lower than $\Delta E_{v-O_M}$ while in U the contrary is true. For C and $O_M$, although CNs don't change on reconstructing, hybridization between carbon and oxygen states within each $CO_3^{2-}$ group (Fig 2.a) renders a downshift of the corresponding electronic states in RR (Fig. 2f). On the other hand, Ca-O bond lengths in RU mainly increase (as discussed above). Accordingly, electron energies associated with C and O in RU increase (Fig. 2e,f).

We have also looked into surface Ca atoms, each of which forms two additional bonds with oxygen on reconstructing. $\Delta E_{v-Ca}$ of the reconstructed surface is 0.15 eV lower than that of the unreconstructed surface. The less significant variation in Ca electron energy may be related to the lower lying of Ca electron states (Fig. 2a), which are therefore less sensitive to chemical environment change. In summary, electron energies of all C and O atoms decrease in RR, as a result of CN increase and hybridization between oxygen and carbon states. Ca electrons also shift down due to CN increase although the change is relatively small. On the contrary, electron energies of all C and O atoms shift up in RU caused by the elongation of Ca-O bonds. Therefore, coordination number increase is the key in stabilizing the reconstructed surface, which decrease electron energies of surface Ca atoms and electron energies of C/O atoms in RR.

**Dynamic stability of calcite (104)**

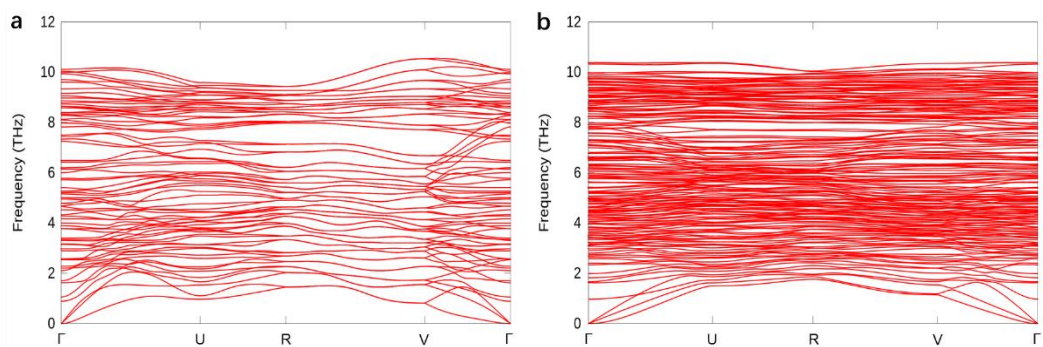

**Fig. 3 | Vibrational properties of calcite (104). a** Phonon spectrum of the unreconstructed surface. **b** Phonon spectrum of the reconstructed surface. Special points in the two-dimensional Brillouin zone: Γ(0,0), U(0.5,0), R(0.5,0.5) and V(0, 0.5).

By performing MD simulations Rohl et al. showed there was an imaginary phonon mode for the unreconstructed surface and suggested that the (2x1) reconstruction was a spontaneous process[22]. However, in a more recent study Magdans et al. reported the unreconstructed (104) surface under both dry and humid atmospheric conditions with GIXRD[17]. This signals the stability of the unreconstructed surface and therefore one would expect an energy barrier separating the unreconstructed and the reconstructed states. To address this problem, we have calculated phonon spectrums for both unreconstructed and reconstructed (104) as shown in Fig. 3. It is apparent that there are more bands in the phonon spectrum of the reconstructed surface, as a result of the symmetry reduction. Rohl reported that the imaginary phonon mode of the unreconstructed surface located at (0.5,0) in the Brillouin zone[22], which is the U point in Fig. 3. However, our DFT simulations show that although U is a local minimum of the lowest phonon band for the unreconstructed surface, the corresponding frequency is positive (~0.9 THz). In both panels of Fig. 3 no imaginary phonon mode exists. Consequently, our study presents a picture that both the unreconstructed and reconstructed surfaces are dynamically stable. We have also calculated phonon spectrum of the unreconstructed surface applying PBE functional and the DFT-D3 method with Becke-Johnson damping[36, 37] (illustrated in Supplementary Fig. 3). The results show that there is no imaginary phonon mode associated with the unreconstructed surface, irrespective the method applied within the DFT framework. Therefore, our DFT results supports Magdans's work which reported the unreconstructed (104) surface under both dry and humid conditions[17].

**Probing the possible energy barrier during reconstructing**

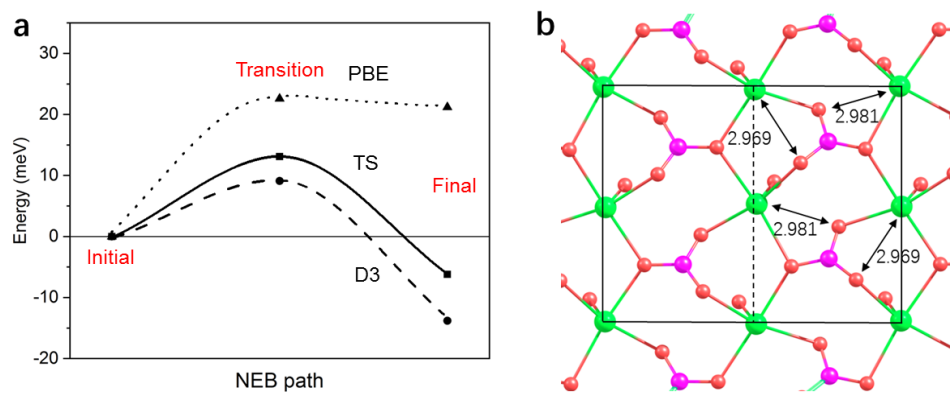

**Fig. 4 | Probing the possible energy barrier during calcite (104) reconstructing. a** Minimum energy path calculated by the CI-NEB method. 'Initial', 'Transition' and 'Final' represent the initial state (unreconstructed surface), the transition state, and the final state (reconstructed surface), respectively. **b** Atomic structure of TS.

We have also applied the CI-NEB method[33] to probe the possible energy barrier. As Fig. 4a shows, there appears to be a single energy barrier between the two states. Fig. 4b shows that in the transition state carbonate groups rotate about $10^0$ in the right half of the (2x1) unit cell, compared with $18^0$ in the final reconstructed state. The two Ca-O distances corresponding to the two 'newly formed' Ca-O bonds when fully reconstructed are 2.969 and 2.981 Å, respectively. These values are about 0.6 Å greater than the nearest Ca-O distance in bulk (2.368 Å) and are about 0.5 Å smaller than the NN Ca-O distance (3.481 Å). Therefore, the transition state can be understood as a state where these two additional Ca-O bonds are about to be formed. The activation energy ($E_a$) is 14 meV per (2x1) unit cell for the formation of the reconstructed surface. Conversely, $E_a$ for the transition from a reconstructed surface to an unreconstructed one is 20 meV per unit cell. Considering that the calculated activation energy is rather small, we have also applied the PEB functional and the DFT-D3 method with Becke-Johnson damping[36, 37] to check this. In line with literature[11, 24], while both DFT methods with van der Waals interactions predict that the reconstructed surface is slightly favored, the semilocal PEB functional predicts that the unreconstructed one is favored. Nevertheless, Fig. 4a shows that there is a non-vanishing but low energy barrier between the unreconstructed and the reconstructed surface, irrespective of the DFT method used.

In summary, by applying DFT methods with van der Waals corrections, we unambiguously show that the calcite (104)-(2x1) reconstruction is driven by the demands of surface atoms to increase the coordination numbers. At the surface four additional Ca-O bonds are formed per (2x1) unit cell on reconstructing as a result of carbonate group rotation. Within the surface layer, electron energies associated with carbonate groups in RR and Ca atoms are effectively lowered due to new bonds forming. In contrast to Rohl's work[22] which performed MD simulations, our DFT phonon spectrums indicate both the unreconstructed and the reconstructed surfaces are stable. The low energy barrier obtained by CI-NEB method suggests that the detected surface structure should be critically dependent on the experimental conditions. Very recently, it has been shown that calcite (2x1) reconstruction can have a decisive impact on calcite adsorption properties[11, 21, 38]. We believe our findings can significantly advance the understanding of surface-related phenomena of calcite.

## Methods
### DFT calculations
All calculations are performed via the Vienna ab-initio simulation package (VASP)[39] Projector-augmented wave (PAW) potentials[40] with the Tkatchenko–Scheffler method with iterative Hirshfeld partitioning[41]. Plane-wave cutoff energy is set to 500 eV and the convergence of energy is $10^{-6}$ eV. 5x8x5 and 2x3x1 Monkhorst–Pack k-point meshes are set to calculate calcite bulk and (104), respectively. Atomic and electronic structure analyses are based on 7-layer fully-optimized structures, although we have found a 4-layer model is thick enough to obtain the converged surface energy (in line with Rahe's work[11]). Phonon analyses and CI-NEB calculations are performed via the 4-layer model (including 12 atomic layers). Phonon spectrums are calculated using the finite difference approach. In NEB calculations we include four images between the initial state and the final state to start the simulations. We have also tested including six images, with the results essentially the same.